# Inhibitory neuristor based on metal-to-insulator transition

*Victor Palin[1], Akash Agnihotri[2], Nareg Ghazikhanian[1], Matthew Frame[3], Yayoi Takamura[3], Ivan K. Schuller[1], Pavel Salev[2]*

[1]*Department of Physics, University of California San Diego*
[2]*Department of Physics & Astronomy, University of Denver*
[3]*Department of Materials Science & Engineering, University of California Davis*

**Mimicking the collective excitatory and inhibitory behaviors of biological neurons remains a critical challenge in the development of neuromorphic computing systems that rival the complexity and performance of the human brain. Volatile high-to-low resistance switching in *insulator-to-metal* transition (IMT) materials produces an abrupt increase in current flow, resembling neuronal excitation. This electrical excitation enables IMT materials to be driven into a neuron-like spiking self-oscillation regime using simple RC circuits. Here, we report a new type of self-oscillation dynamics that occurs in the opposite class of *metal-to-insulator* transition (MIT) materials. Electrical triggering of the MIT suppresses current flow, resembling neuronal inhibition. Using a prototypical MIT material, we experimentally demonstrate inhibitory-like self-oscillations in two-terminal switching devices incorporated into a simple RL circuit. Our results show robust ~0.1 – 1 MHz electric current oscillations with minimal cycle-to-cycle variation, which can be controlled by varying the applied DC voltage, temperature, and inductance. This work demonstrates a new type of inhibitory MIT-based artificial neuron that can complement the excitatory functionalities of IMT-based neuristors in biologically plausible neuromorphic systems.**

**Introduction**

The close resemblance between the electrically driven self-oscillations in insulator-metal transition (IMT) materials and excitatory spiking in biological neurons has prompted extensive research efforts to understand and control self-oscillation behavior in IMT switching devices [1–7], explore synchronization and emergent collective phenomena in interacting oscillators [8–12], and implement hardware-level spiking neural networks [13–17]. The key property enabling the IMT self-oscillations is the volatile switching from high to low resistance produced by the electrical triggering of the phase transition [18,19]. When an IMT device is incorporated into an RC circuit, neither high- nor low-resistance states are stable at certain DC biases. As a result, the device transiently switches into a metallic state and then relaxes back, which produces periodic current surges, i.e., spikes. The simplicity of the IMT-based RC oscillator circuits enables high scalability and high energy efficiency, making these oscillators promising for neuromorphic hardware applications. Self-oscillating behavior has been observed in a wide range of IMT materials with vastly different phase transition mechanisms and electronic, structural, and magnetic properties, including $NbO_2$, $VO_2$, $V_2O_3$, $V_3O_5$, $NdNiO_3$, 1T-$TaS_2$, etc. [2,20–28]. At this point it is reasonable to assume that any material exhibiting an IMT-based volatile resistive switching can also be driven into the electrical self-oscillation regime.

Metal-insulator transition (MIT) materials, such as many members of the rare-earth manganite family, exhibit opposite properties compared to the IMT materials. In MIT systems, the phase transition occurs from a metallic ground state to an insulating state, resulting in a large resistivity increase across the critical temperature [29,30]. The MIT can also be triggered electrically, which produces a volatile low-to-high resistance switching [31,32]. In contrast to the percolating metallic-phase longitudinal filaments in IMT switches [33–35], the MIT switching occurs by the formation of an insulating-phase transverse barrier that splits the metallic matrix and attenuates current flow [31,36]. Electrical triggering of the MIT can also be used to control local magnetic properties [37–40], adding another layer of practical functionality to MIT switches.

Even though electrical switching has been observed in many MIT materials [31,32,41–44], the possibility of driving MIT switches into a neuron-like self-oscillating regime has remained unexplored until now. Here, we demonstrate that fast electrical self-oscillations can be induced by applying a DC voltage to MIT switching devices integrated into a simple RL circuit. These self-oscillations are highly robust, emerging across a wide range of applied voltages, temperatures, and inductances, and in multiple samples and devices of different dimensions, suggesting that the self-oscillation behavior is a general phenomenon that likely to be found in a variety of MIT materials. While momentary switching into a low-resistance state generates sharp current spikes in IMT oscillators, we observed periodic current dips due to switching into an insulating state in our MIT oscillators. These distinct current-spike and current-dip operations in IMT and MIT switches, respectively, provide two complementary electrical oscillatory dynamics that could enable the design of artificial neural networks capable of mimicking the collective excitatory and inhibitory behaviors of biological systems.

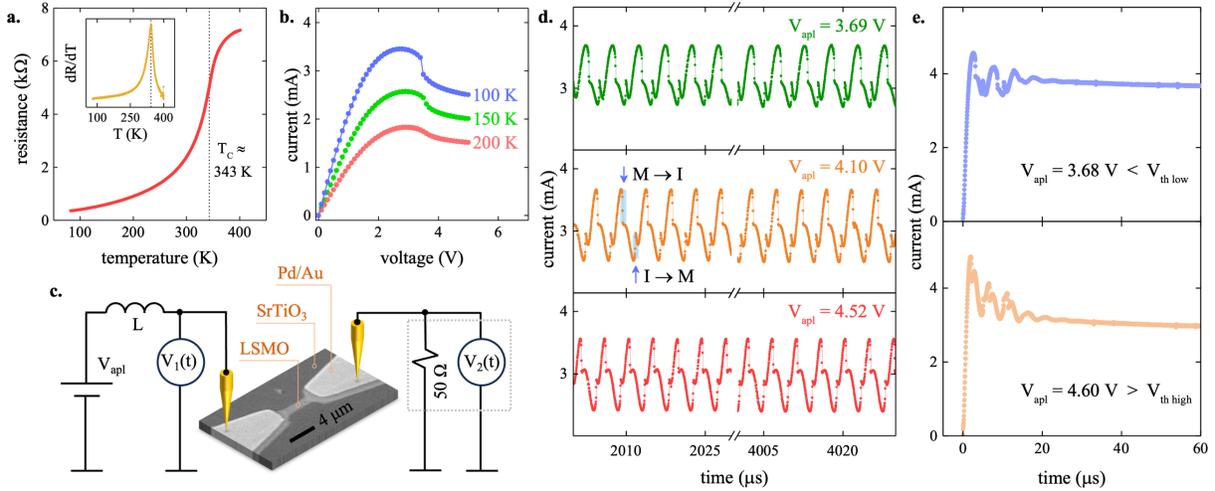

**Fig. 1. a.** Resistance-temperature dependence of an LSMO device exhibiting the MIT. Inset shows the resistance derivative in which the peak corresponds to the transition temperature of 343 K. **b.** I-V characteristics recorded at several temperatures showing high-to-low resistance switching and negative differential resistance (a part of the curve where $dI/dV < 0$). **c.** RL electrical circuit used in the oscillation measurements. The circuit is powered by a DC voltage source ($V_{apl}$). Two oscilloscope channels, $V_1(t)$ and $V_2(t)$, were used to monitor electrical dynamics in the circuit. **d.** Current time traces recorded at several applied DC voltages at 100 K revealing persistent electrical oscillations. The measurements were performed using a 1 mH inductor. **e.** Current time traces recorded at applied DC voltages below (top panel) and above (bottom panel) the applied DC voltage window of stable oscillations. All measurements were performed using a 4×1 μm² LSMO device.

**Results**

To search for electrical self-oscillations in MIT switches, we used planar two-terminal devices patterned on a 20-nm-thick La$_{0.7}$Sr$_{0.3}$MnO$_3$ (LSMO) film epitaxially grown on a (001)-oriented SrTiO$_3$ substrate. Fig. 1a shows the resistance-temperature dependence of a 4×1 μm² device. The measurements revealed a steep increase in resistance above 300 K, which corresponds to the low-temperature ferromagnetic metal to high-temperature paramagnetic insulator phase transition in LSMO [30]. The resistance derivative (inset in Fig. 1a) indicates that the MIT temperature in our sample is $T_c \approx 343$ K, as expected for a stoichiometric film [45,46]. The MIT in LSMO can also be driven electrically by applying voltage to the two-terminal device. Electrical triggering of the MIT results in highly nonlinear I-V characteristics that exhibit an N-type negative differential resistance, indicative of the low-to-high resistance switching (Fig. 1b) [31]. We note that this electrical switching is volatile. The high-resistance state is induced when the applied voltage triggers the MIT, and the device automatically resets to the original low-resistance state when voltage is turned off. Overall, our sample exhibited phase transition and electrical switching characteristics typically found in LSMO thin film devices [31,37,38].

Self-oscillations in IMT materials such as VO$_2$, NbO$_2$, NdNiO$_3$, etc., are commonly observed when the switching device is incorporated into an RC circuit [2,20–27]. In an RC circuit, a series resistor and the IMT device form a voltage divider that regulates the applied voltage distribution, while the discharge of the parallel capacitor helps prevent the IMT device from settling into a stable metal-filament state after the switching. An RC-type circuit, however, is not suitable for inducing electrical self-oscillations in MIT materials. In such a circuit, the MIT device focuses a large fraction of the applied voltage after switching into the insulating phase, which stabilizes the switching. Our experiments with MIT devices incorporated into RC circuits showed no oscillatory behavior.

We propose a new RL-type self-oscillating circuit (Fig. 1c). In this circuit, the MIT device is connected in series with an inductor, and the circuit is powered by a DC voltage source. Two oscilloscope channels are used to monitor the voltage across the MIT device and the circuit's current by measuring the voltage across a 50 Ω shunt resistor. The operation of the MIT-based RL circuit can be understood as a dynamic interplay between the abrupt resistance switching of the MIT device and the inductor's tendency to oppose rapid changes in current. Initially, the MIT device is in the metallic state. An applied DC voltage triggers the MIT, which occurs by the formation of a transverse insulating barrier that impedes the current flow [31]. The purpose of the series inductor is to prevent the MIT device from settling into a stable high-resistance state after switching. Because the insulating barrier formation causes a transient current decrease, the inductor generates an EMF, $\mathcal{E} = -L(dI/dt)$, that counteracts this decrease. The voltage across the MIT device becomes a sum of the applied voltage and inductor's EMF. As a result, the device is driven deep into the insulating state, which cannot be sustained by the DC voltage alone. When the MIT device begins to relax from this insulating state overshoot and the circuit current increases, the inductor's EMF counteracts the increase in current. The voltage across the MIT device becomes the difference between the applied voltage and the inductor's EMF. This effective reduction in voltage causes the device to reset to the initial metallic state. Thus, during each oscillation cycle, the MIT device switches from metal to insulator and then relaxes back from insulator to metal.

We observed persistent self-oscillation electrical dynamics under the application of DC voltage when the LSMO switching device is incorporated into an RL circuit. Fig. 1d shows several examples of the current time traces recorded using DC voltages in the 3.69 – 4.52 V range (the oscillation window in this experiment). These measurements were performed using a 4×1 μm$^2$ device at 100 K. The current time traces exhibit clear oscillations with an amplitude of ~1 mA and a period of a few microseconds. The oscillating current curves have nontrivial shapes, including regions of gradual change and points of abrupt jumps corresponding to MIT switching (highlighted by blue arrows in Fig. 1d). The oscillations remain stable over long periods: as long as the applied DC voltage is maintained, the LSMO device continues to exhibit identical oscillations. The oscillations also have extremely low cycle-to-cycle variability. The standard deviations of the oscillation period and amplitude are typically below 0.5%, approaching the time and voltage resolution limits of our oscilloscope setup. This low cycle-to-cycle variability in MIT-based oscillators contrasts sharply with the pronounced stochasticity commonly observed in IMT-based oscillators [9,10,47–50]. While the physical origin of this extremely low oscillation variability is unknown, several factors may contribute, including the 2$^{nd}$ order nature of the MIT in LSMO, the absence of an abrupt, 1$^{st}$ order structural transition during the MIT, the relatively low metal–insulator resistance contrast, and the uniform current distribution when the device is in the metallic state [31,36,51]. The self-oscillations in LSMO occur within a specific applied DC voltage window. When the applied voltage is outside this window, a few rapidly decaying oscillations can be observed before the current settles to a constant value (Fig. 1e). Overall, our measurements demonstrate that it is possible to drive MIT materials, such as LSMO, into an electrical self-oscillation regime by applying a DC voltage in a simple RL circuit.

We found evidence that during each oscillation cycle, the LSMO device first undergoes metal-to-insulator (M→I) switching and then the reverse insulator-to-metal (I→M) transition. Fig. 2a shows the time dependence of the LSMO device resistance (top panel), the voltage drop across the device (middle panel), and the power dissipated in the device (bottom panel). These measurements were performed on a 4×1 μm$^2$ device at 80 K. Similar to the current time trace in Fig. 1d, the resistance, voltage, and dissipated power exhibit periodic oscillations. Interestingly, the instantaneous voltage across the device can exceed the applied DC voltage (4.96 V vs. 3.75 V in this example). This voltage overshoot is due to the inductive response of the circuit, i.e., the inductor's EMF and the DC voltage add together. The resistance peaks and dips lag those of the voltage and power by ~0.3 μs (highlighted by the vertical dashed lines in Fig. 2a). Because resistance directly reflects the electronic state of LSMO (i.e., the metal and insulator phase fractions), the time lag between resistance and voltage (power) suggests that the device is not in equilibrium with respect to the instantaneous electric field (Joule heating) in the self-oscillation regime. Consequently, the resistance exhibits a stable closed-loop hysteresis in the self-oscillation regime, as shown in the phase-space plots in Fig. 2b. We note that these hysteresis plots include hundreds of overlaid cycles that are virtually indistinguishable from one another. Under all experimental conditions (applied voltages and series inductances), the resistance–voltage hysteresis exhibits a characteristic two-lobe shape, with the lobes separated by abrupt resistance jumps. The abrupt resistance increase can be attributed to M→I switching (i.e., insulating barrier formation), while the abrupt resistance decrease corresponds to I→M switching (i.e., collapse of the barrier and recovery of the metallic state). We note that the maximum resistance during oscillations (~2.7 kΩ, Fig. 2b) is several times smaller than the resistance of the insulating state in the resistance–temperature dependence (~7.2 kΩ, Fig. 1a). This apparent discrepancy can be understood by considering that electrical MIT switching occurs by the formation of a localized insulating barrier, i.e., only part of the device undergoes the MIT [31], whereas in temperature-dependent measurements the phase transition occurs throughout the entire material volume. We therefore conclude that the LSMO device undergoes MIT switching during each oscillation cycle, first M→I and then the reverse I→M transition.

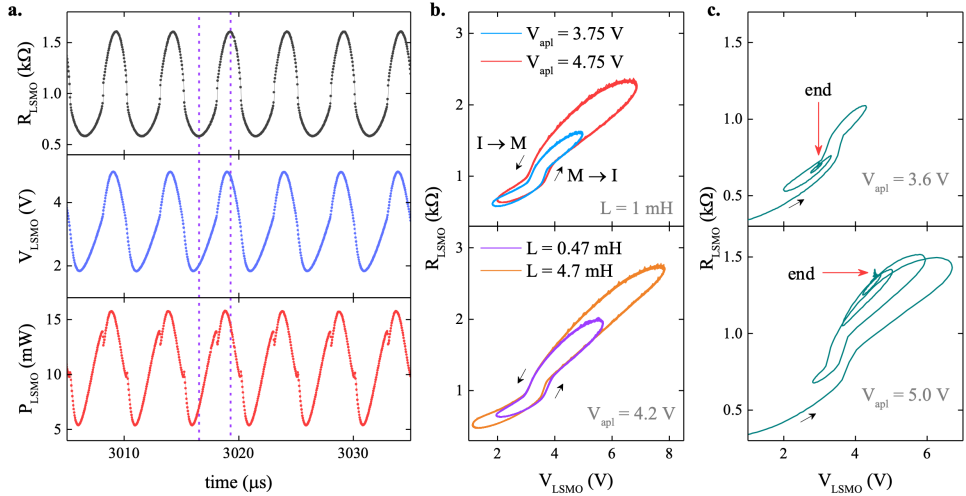

**Fig. 2. a.** Oscillations of the LSMO device resistance (top panel), the voltage across the device (middle panel), and the power dissipation in the device (bottom panel) under the application of 3.75 V DC voltage. Resistance oscillations lag the voltage and power oscillations by ~0.3 μs. **b.** Phase-space plots showing the "device resistance" vs. "voltage drop across the device" during stable oscillations. Data recorded using different applied DC voltages (top panel) and different inductances (bottom panel) are shown, exhibiting similar closed-loop, two-lobe hysteresis. Hundreds of virtually indistinguishable loops are overlaid in the plots. **c.** Phase-space plots demonstrating unstable oscillations, i.e., electrical dynamics when the applied voltage is either below (top panel) or above (bottom panel) the DC voltage window of stable oscillations. After a few oscillations, the LSMO device quickly settles either into a metallic or an insulating state. All measurements were performed using a 4×1 μm$^2$ device at 80 K.

Undergoing M→I and I→M switching events during each cycle is a necessary condition for maintaining stable oscillations in our LSMO devices. Fig. 2c shows the "device resistance" – "device voltage" phase-space plots corresponding to electrical dynamics at applied voltages just outside the oscillation window. In these plots, one can observe a few initial M→I and I→M switching events (the two-lobe hysteresis features). However, as soon as the device stops undergoing the MIT switching, the oscillations rapidly cease and LSMO settles in a stable low- or high-resistance state. We note that LSMO exhibits a nonlinear resistance-temperature dependence at temperatures below and above the MIT (Fig. 1a). The resistance nonlinearity in either the low-temperature metallic state or the high-temperature insulating state appears to be insufficient to sustain persistent self-oscillation dynamics in the RL circuit. Therefore, if the applied DC voltage is too low to induce MIT switching in LSMO, or if it is too high and causes the device to remain in the switched state, stable electrical self-oscillations are not induced.

We observed stable oscillations over a wide range of experimental control parameters (voltage, temperature, inductance) and in devices with different dimensions. Fig. 3a shows the oscillation frequency and amplitude dependence on applied voltage and temperature. In a 10×2 μm$^2$ device at 80 K, the applied DC voltage window for oscillations is relatively wide, ~5.3 – 9.9 V, and the frequency can be tuned within a ~41 kHz range by increasing the DC voltage. At higher temperatures, the DC voltage oscillation window shrinks (inset in Fig. 3a, top panel), while the oscillation frequency increases considerably. The oscillation amplitude (Fig. 3a, bottom panel) exhibits a nonmonotonic dependence on DC voltage, and the amplitude decreases as temperature increases. This amplitude decrease is likely due to the reduced resistance contrast between the metallic and insulating phases at higher temperatures.

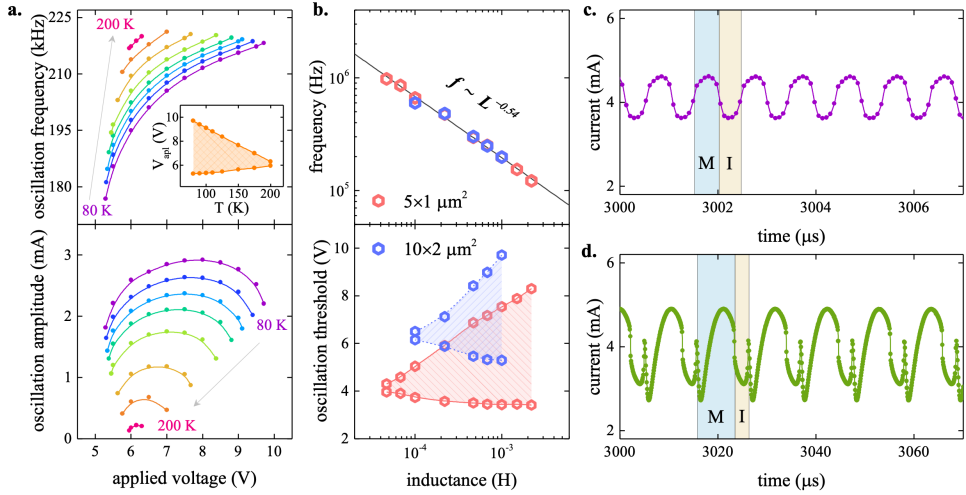

**Fig. 3. a.** Temperature and voltage dependence of the oscillation frequency (top panel) and amplitude (bottom panel). The measurements were performed using a 10×2 μm² device. The inset in the top panel shows the applied DC voltage window of stable oscillations (shaded area) as a function of temperature. The oscillations can be induced over a wide range of voltages and temperatures. **b.** Inductance dependence of the oscillation frequency and the applied DC voltage window of oscillations (shaded areas). The measurements were performed at 80 K. Oscillations can be induced using a wide range of inductances. Multiple devices of different dimensions show similar oscillation behavior. **c, d.** Current time traces corresponding to the fastest (c) and slowest (d) oscillations in a 5×1 μm² device at 80 K recorded using 33 μH and 2.2 mH series inductances, respectively. Shaded regions highlight the periods when the LSMO device is either in the metallic (M) or insulating (I) state.

Fig. 3b shows how the oscillation frequency and the oscillation DC voltage window depend on the series inductance. These measurements were performed using 5×1 μm² and 10×2 μm² devices at 80 K. In general, the lower the inductance, the faster the oscillations, but the narrower the oscillation window. The oscillation window also depends on the device dimensions. It is possible, however, to induce stable oscillations using the same applied DC voltage in devices whose LSMO volume differs by at least a factor of 4 (the overlap region in the bottom panel plot of Fig. 3b). We note that there is a minimum series inductance at which driving LSMO devices into the oscillation regime is possible (47 μH for 5×1 μm² and 100 μH for 10×2 μm² devices in this experiment). While we found no upper bound for the series inductance required to induce oscillations, we observed that using inductances above a few mH results in device degradation because of the large instantaneous voltage associated with the inductor's EMF.

Interestingly, we observed a power-law scaling of the oscillation frequency with inductance, $f \sim L^{-0.54}$ (Fig. 3b). The scaling exponent, 0.54, deviates significantly from 1, which indicates that the oscillation dynamics are not governed purely by the RL time constant, $\tau = L/R$. The dynamics of electrically induced phase transition in LSMO likely play a major role in defining the oscillation properties. Overall, we were able to induce persistent electrical self-oscillations at different temperatures, applied voltages, and inductances in multiple devices and in multiple LSMO samples, suggesting that the observed oscillation phenomena may be a general property of MIT switching materials.

The oscillation frequency of the LSMO device is highly tunable and can reach relatively high values, ~0.1 – 1 MHz in the inductance dependent measurements (Fig. 3b, top panel). At high frequencies (Fig. 3c), the current oscillation time trace resembles a sine wave, and LSMO spends comparable periods in the metallic and insulating states (identified as the current peaks and dips): ~0.54 μs vs. ~0.42 μs, respectively (shaded regions in Fig. 3c). The oscillation amplitude at high frequencies is also relatively low, ~1 mA, indicating that only a small volume of the device undergoes oscillatory MIT switching during each cycle. At low frequencies (Fig. 3d), the current oscillation time trace differs drastically from a simple sine wave, and the oscillation amplitude, ~2.2 mA, is larger than that of the high-frequency oscillations, indicating that a larger volume of LSMO undergoes switching during each cycle. By identifying abrupt current jumps in the time traces, we outlined the periods when the LSMO device is either in the metallic or insulating state (shaded regions in Fig. 3d). In contrast to the high-frequency oscillations, the LSMO device spends most of time, 7.84 μs, in the metallic state, and switches only briefly, 2.72 μs, into the insulating state during each low-frequency oscillation cycle. Because in both high- and low-frequency oscillation regimes the LSMO device spends more time in the metallic state, the voltage buildup on the device to trigger M→I switching is likely the limiting factor controlling the oscillation frequency. We note that we observed a maximum oscillation frequency of ~1 MHz in a 5×1 μm² microscale device. In IMT oscillators (e.g., based on VO$_2$), MHz frequencies can be achieved only in nanosize devices, while the microscale devices typically oscillate in a 1 – 100 kHz range [23]. It is possible that oscillatory dynamics with frequencies significantly higher than 1 MHz can be achieved by reducing the LSMO device dimensions to the nanoscale.

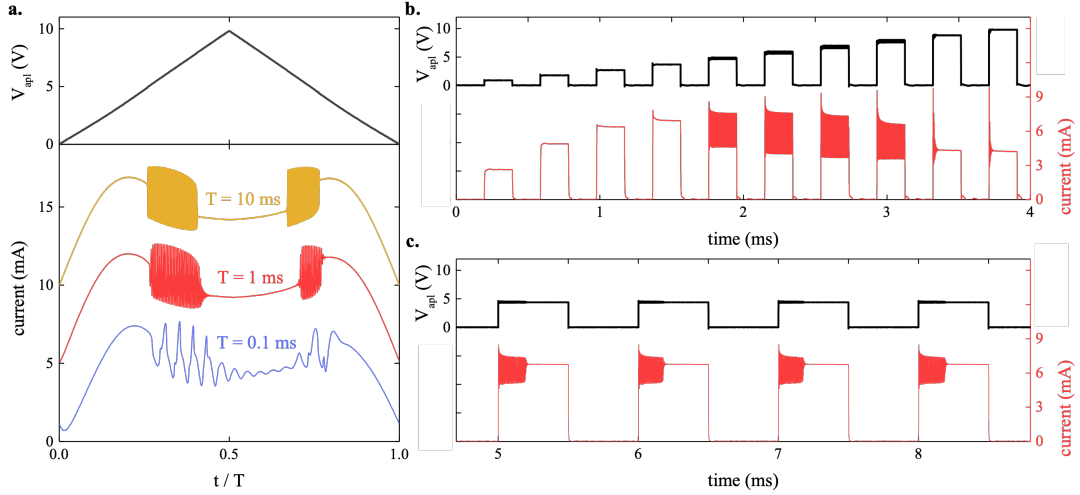

**Fig. 4. a.** Electrical self-oscillations during the application of a voltage ramp. The horizontal time axis is normalized by the ramp period *T*. The top panel shows the measured applied voltage during a 10-ms-period ramp (1-ms and 0.1-ms ramps look identical with the normalized time axis). The bottom panel shows the current through the LSMO device recorded using 10-ms (yellow), 1-ms (red), and 0.1-ms (blue) ramps. The 1-ms and 10-ms current traces are vertically offset by 5 mA and 10 mA, respectively. In all three current traces, conducting, oscillating inhibition, and persistent inhibition operation modes can be observed. **b.** Electrical self-oscillations (red line) induced by progressively increasing (in steps of 1 V) applied voltage pulses (black line). Small amplitude pulses are transmitted, medium amplitude pulses induce inhibitory oscillations, large amplitude pulses are attenuated by the LSMO switching into a persistent insulating state. **c.** Oscillation adaptation (red line) when the applied voltage (black line) is on the verge of the stable oscillation window. The LSMO device exhibits oscillations that abruptly halt at ~0.2 ms after stimulus application.

To highlight the current-inhibitory character of the MIT-based oscillator, we performed experiments using time dependent applied voltage. Fig. 4a shows the emergence of self-oscillations under applied triangular voltage ramps with varying periods. These measurements were performed using an 8×2 μm² device at 80 K. At first, when the applied voltage increases, the current also increases rapidly because the LSMO device is in the metallic state. As the applied voltage becomes sufficiently high to bring the device close to the MIT, a negative differential resistance region develops, which limits the current flow. At higher applied voltages, oscillations set in. Each oscillation momentarily disrupts the current because the device briefly switches into the high-resistance state. Finally, as the applied voltage increases further, the device switches into a persistent insulating state, and the current flow is inhibited as the voltage continues to increase. Even at a short period of 0.1 ms for the applied voltage ramp, we can observe all three operation modes: metallic conductance, inhibitory oscillations, and persistent current inhibition. These measurements also demonstrate the volatility of the MIT switching in LSMO: oscillations and persistent switching are induced when the applied voltage is high, and the device resets into its original metallic state once the applied voltage is reduced.

Fig. 4b shows the oscillator response to 0.2-ms square voltage pulses with amplitudes increasing in 1 V steps. The first four pulses (1 – 4 V) are too small to trigger the MIT, therefore, the pulses are transmitted without inhibition, i.e. the device shows metallic conduction operation. The next four pulses (5 – 8 V) drive the LSMO into the oscillation regime, during which the current flow is momentarily disrupted due to the periodic MIT switching, i.e., the device shows oscillatory inhibition operation. The last two pulses (9 and 10 V) switch the LSMO into the insulating state, resulting in strong current attenuation, i.e., the device shows persistent inhibition operation.

At applied voltages near the stable oscillation window, the LSMO oscillator exhibits interesting adaptation-like behavior. Fig. 4c shows the oscillator circuit response to applied square voltage pulses of 4.7 V amplitude and 0.5 ms duration. These applied voltage pulses drive the LSMO into the oscillation regime. The oscillations, however, abruptly halt after ~0.2 ms, even though the applied voltage remains unchanged. This oscillation-halting behavior is highly reproducible – the same applied voltage stimulus always leads to the same device response. Because there are 34 well-defined large-amplitude oscillations during this ~0.2 ms period, the abrupt oscillation halting cannot be attributed to trivial electrical decay in the RL circuit. A possible explanation of the oscillation halting behavior is the local heating of the substrate due to the current flowing in the MIT device. The applied DC voltage required to initiate the oscillations gradually increases with increasing temperature (Fig. 3a). After a few current oscillations in LSMO, Joule heating may locally raise the temperature of the device vicinity, causing the constant applied voltage to fall outside the stable oscillation window. This observed oscillation-halting behavior in LSMO resembles biological neuron adaptation, where maintaining a persistent stimulus eventually halts neuronal spiking. Our results, therefore, point to an exciting opportunity to mimic advanced neuronal functions in MIT-based oscillators for the hardware implementation of bio-realistic neural networks.

**Discussion**

While electrical self-oscillation phenomena have already been firmly established in IMT materials, this work introduces a new class of self-oscillators based on the opposite type of phase transition in MIT materials. Here, we demonstrate the emergence of electrical oscillations in a prototypical MIT material, LSMO. Because we observed oscillations over a broad range of control parameters and device dimensions, it is natural to expect that electrical oscillations under DC voltage application can be induced in different MIT materials. Further experiments, e.g., in other members of the rare-earth manganite family [29,30], are needed to gain the general understanding of the self-oscillation phenomena in MIT-based circuits and to identify the most promising materials for practical applications.

Understanding spatio-temporal dynamics could be key to maximizing the self-oscillation frequency in MIT materials. Our measurements showed that the resistance oscillation amplitude in LSMO is several times smaller than the full metal-insulator contrast in the resistance-temperature dependence (Fig. 2a vs. Fig. 1a), indicating that only a small region of the device undergoes switching during each cycle. Quantifying the size and dynamics of the switched region during self-oscillations requires a probe with high temporal and spatial resolution that is sensitive to either electronic (metal-insulator), magnetic (ferromagnetic-paramagnetic), or structural (lattice expansion) contrast in LSMO switching devices [31,36,40]. Minimizing the switching region size, for instance, by nanoscale confinement, may lead to faster oscillations and lower voltage/power operation. Furthermore, MIT device miniaturization can alleviate the need for a bulky external inductor, similar to nanoscale IMT switches that can oscillate without connecting to an external parallel capacitor by relying solely on the circuit's parasitic capacitance [4,23,50].

**Summary**

In summary, we have experimentally demonstrated electrical self-oscillation dynamics under the application of DC voltage in a metal-to-insulator transition material, $La_{0.7}Sr_{0.3}MnO_3$, using a simple RL circuit. The oscillation characteristics of the MIT-based RL circuit, including frequency, amplitude, and waveform shape, can be continuously tuned by adjusting the applied voltage, series inductance, or temperature. We were able to readily induce oscillatory dynamics in multiple devices and samples, which suggests that the self-oscillation behavior is likely not unique to LSMO switches and can potentially be engineered into other materials that transition between a low-temperature metallic phase and a high-temperature insulating phase. During each oscillation cycle, the MIT material briefly switches into a high-resistance state, which disrupts the current flow in the circuit. This current disruption behavior is in contrast with self-oscillations in IMT materials, in which a transient switching into a low-resistance state results in the generation of a current spike. Together, the IMT- and MIT-based oscillators provide a wide range of dynamic electrical operation regimes, greatly expanding the design space for realistic mimicking of excitatory and inhibitory behaviors of biological neuronal systems.

**Methods**

Sample preparation. 20-nm-thick $La_{0.7}Sr_{0.3}MnO_3$ films were epitaxially grown on (001)-oriented $SrTiO_3$ substrates by pulsed laser deposition using a laser fluence of 0.6 J/cm$^2$ and a repetition rate of 1 Hz. The substrates temperature was 700 °C and the oxygen pressure was 300 mTorr during the film growth. After deposition, the films were gradually cooled to room temperature in 300 Torr oxygen pressure. (20 nm Pd)/(20 nm Au) low-contact-resistance electrodes and (20 nm Ti)/(100 nm Au) wire bonding pads were fabricated using standard photolithography and e-beam evaporation. Isolated switching devices were patterned using Ar ion milling.

Electrical measurements. The measurements were performed in a liquid nitrogen cooled Lakeshore Cryotronics TTPX probe station. Quasi static measurements (resistance-temperature dependence, I-V characteristics) were performed using a Keithley 2450 source meter. Oscillation dynamics measurements were performed using a Tektronix AFG3252C function generator and a Tektronix MSO54 oscilloscope. Commercial inductors were used for assembling the RL oscillator circuit.

**Acknowledgement**

Work at DU was supported by the U.S. Department of Energy, Office of Science, Basic Energy Sciences, under Award # DE-SC0026129. Work at UCSD and UCD was supported as part of the Quantum Materials for Energy Efficient Neuromorphic Computing (Q-MEEN-C), an Energy Frontier Research Center funded by the U.S. Department of Energy, Office of Science, Basic Energy Sciences under Award # DE-SC0019273.